\documentclass[12pt,a4paper]{article}
\usepackage{psfrag}
\usepackage{graphics}
\usepackage{amssymb,epsfig,amsmath,euscript,array}

\makeatletter
\@addtoreset{equation}{section}
\makeatother

\newcounter{multieqs}

\newcommand{\be}{\begin{equation}}
\newcommand{\ee}{\end{equation}}

\newcommand{\bm}[1]{\mbox{\boldmath $#1$}}

\def\bd{\begin{document}}
\def\ed{\end{document}}
\def\nn{\nonumber}
\def\bea{\begin{eqnarray}}
\def\eea{\end{eqnarray}}
\let\bm=\bibitem
\let\la=\label

\newcommand{\EQ}[1]{\begin{equation} #1 \end{equation}}
\newcommand{\AL}[1]{\begin{subequations}\begin{align} #1 \end{align}\end{subequations}}
\newcommand{\SP}[1]{\begin{equation}\begin{split} #1 \end{split}\end{equation}}
\newcommand{\ALAT}[2]{\begin{subequations}\begin{alignat}{#1} #2 \end{alignat}\end{subequations}}
\def\beqa{\begin{eqnarray}}
\def\eeqa{\end{eqnarray}}
\def\beq{\begin{equation}}
\def\eeq{\end{equation}}

\def\hf{{\textstyle{1\over2}}}
\def\wbar{\bar w}
\def\mubar{\bar\mu}
\def\abar{\bar a}
\def\sigmabar{\bar\sigma}
\def\etabar{\bar\eta}
\def\zetabar{\bar\zeta}
\def\mubar{\bar\mu}
\def\nubar{\bar\nu}
\def\N{{\cal N}}
\def\sst{\scriptscriptstyle}
\def\thetabar{\bar\theta}
\def\Tr{{\rm Tr}}
\def\one{\mbox{1 \kern-.59em {\rm l}}}
 \def\Nh{\hat{N}}

\def\a{\alpha}      \def\da{{\dot\alpha}}
\def\b{\beta}       \def\db{{\dot\beta}}
\def\c{\gamma}  \def\G{\Gamma}  \def\cdt{\dot\gamma}
\def\d{\delta}  \def\D{\Delta}  \def\ddt{\dot\delta}
\def\e{\epsilon}        \def\vare{\varepsilon}
\def\f{\phi}    \def\F{\Phi}    \def\vvf{\f}
\def\h{\eta}
\def\k{\kappa}
\def\l{\lambda} \def\L{\Lambda}
\def\m{\mu} \def\n{\nu}
\def\o{\omega}
\def\p{\pi} \def\P{\Pi}
\def\r{\rho}
\def\s{\sigma}  \def\S{\Sigma}
\def\t{\tau}
\def\th{\theta} \def\Th{\Theta} \def\vth{\vartheta}
\def\X{\Xeta}
\def\z{\zeta}

\def\cA{{\cal A}} \def\cB{{\cal B}} \def\cC{{\cal C}}
\def\cD{{\cal D}} \def\cE{{\cal E}} \def\cF{{\cal F}}
\def\cG{{\cal G}} \def\cH{{\cal H}} \def\cI{{\cal I}}
\def\cJ{{\cal J}} \def\cK{{\cal K}} \def\cL{{\cal L}}
\def\cM{{\cal M}} \def\cN{{\cal N}} \def\cO{{\cal O}}
\def\cP{{\cal P}} \def\cQ{{\cal Q}} \def\cR{{\cal R}}
\def\cS{{\cal S}} \def\cT{{\cal T}} \def\cU{{\cal U}}
\def\cV{{\cal V}} \def\cW{{\cal W}} \def\cX{{\cal X}}
\def\cY{{\cal Y}} \def\cZ{{\cal Z}}

\def\ua{\underline{\alpha}}
\def\ub{\underline{\phantom{\alpha}}\!\!\!\beta}
\def\uc{\underline{\phantom{\alpha}}\!\!\!\gamma}
\def\um{\underline{\mu}}
\def\ud{\underline\delta}
\def\ue{\underline\epsilon}
\def\una{\underline a}\def\unA{\underline A}
\def\unb{\underline b}\def\unB{\underline B}
\def\unc{\underline c}\def\unC{\underline C}
\def\und{\underline d}\def\unD{\underline D}
\def\une{\underline e}\def\unE{\underline E}
\def\unf{\underline{\phantom{e}}\!\!\!\! f}\def\unF{\underline F}
\def\unm{\underline m}\def\unM{\underline M}
\def\unn{\underline n}\def\unN{\underline N}
\def\unp{\underline{\phantom{a}}\!\!\! p}\def\unP{\underline P}
\def\unq{\underline{\phantom{a}}\!\!\! q}
\def\unQ{\underline{\phantom{A}}\!\!\!\! Q}
\def\unH{\underline{H}}

\def\As {{A \hspace{-6.4pt} \slash}\;}
\def\bs {{b \hspace{-6.4pt} \slash}\;}
\def\Ds {{D \hspace{-6.4pt} \slash}\;}
\def\ds {{\del \hspace{-6.4pt} \slash}\;}
\def\ss {{\s \hspace{-6.4pt} \slash}\;}
\def\ks {{ k \hspace{-6.4pt} \slash}\;}
\def\ps {{p \hspace{-6.4pt} \slash}\;}
\def\pas {{{p_1} \hspace{-6.4pt} \slash}\;}
\def\pbs {{{p_2} \hspace{-6.4pt} \slash}\;}

\def\Fh{\hat{F}}
\def\Vh{\hat{V}}
\def\Xh{\hat{X}}
\def\ah{\hat{a}}
\def\xh{\hat{x}}
\def\yh{\hat{y}}
\def\ph{\hat{p}}
\def\xih{\hat{\xi}}

\def\psit{\tilde{\psi}}
\def\Psit{\tilde{\Psi}}
\def\tht{\tilde{\th}}

\def\At{\tilde{A}}
\def\Qt{\tilde{Q}}
\def\Rt{\tilde{R}}
\def\Nt{\tilde{N}}

\def\at{\tilde{a}}
\def\st{\tilde{s}}
\def\ft{\tilde{f}}
\def\pt{\tilde{p}}
\def\qt{\tilde{q}}
\def\vt{\tilde{v}}
\def\nt{\tilde{n}}

\def\delb{\bar{\partial}}
\def\bz{\bar{z}}
\def\bD{\bar{D}}
\def\bB{\bar{B}}

\def\bk{{\bf k}}
\def\bl{{\bf l}}
\def\bp{{\bf p}}
\def\bq{{\bf q}}
\def\br{{\bf r}}
\def\bx{{\bf x}}
\def\by{{\bf y}}
\def\bR{{\bf R}}
\def\bV{{\bf V}}

\def\d{\delta}\def\D{\Delta}\def\ddt{\dot\delta}

\def\pa{\partial} \def\del{\partial}
\def\xx{\times}
\def\uno{\mbox{1 \kern-.59em {\rm l}}}

\def\trp{^{\top}}
\def\inv{^{-1}}
\def\dag{{^{\dagger}}}
\def\pr{^{\prime}}

\def\rar{\rightarrow}
\def\lar{\leftarrow}
\def\lrar{\leftrightarrow}

\newcommand{\0}{\,\!}      
\def\one{1\!\!1\,\,}
\def\im{\imath}
\def\jm{\jmath}

\newcommand{\tr}{\mbox{tr}}
\newcommand{\slsh}[1]{/ \!\!\!\! #1}

\def\vac{|0\rangle}
\def\lvac{\langle 0|}

\def\hlf{\frac{1}{2}}
\def\ove#1{\frac{1}{#1}}

\def\Box{\square}
\def\ZZ{\mathbb{Z}}
\def\CC#1{({\bf #1})}
\def\bcomment#1{}
\def\bfhat#1{{\bf \hat{#1}}}
\def\VEV#1{\left\langle #1\right\rangle}
\def\vev#1{\langle{#1}\rangle}

\newcommand{\ex}[1]{{\rm e}^{#1}} \def\ii{{\rm i}}

\def\rr{{\rm r}} \def\rs{{\rm s}}\def\rv{{\rm v}}
\def\ri{{\rm i}}\def\rj{{\rm j}}

\newcommand{\lrbrk}[1]{\left(#1\right)}
\newcommand{\sfrac}[2]{{\textstyle\frac{#1}{#2}}}

\font\mybb=msbm10 at 12pt
\def\bb#1{\hbox{\mybb#1}}

\font\myBB=msbm10 at 18pt
\def\BB#1{\hbox{\myBB#1}}

\begin{document}


\hfill{}

\vspace{20pt}

\begin{center}

{\Large \bf Multi-soft theorems in Gauge Theory  }

{\Large \bf  from MHV Diagrams}

\vspace{30pt}

{\bf George Georgiou }

\medskip

{\small \em
Institute of Nuclear and Particle Physics,
National Center for Scientific Research Demokritos,
15310 Athens, Greece
}

\vspace{10pt}

{\sffamily \tt georgiou@inp.demokritos.gr }

\vspace{30pt}

{\bf Abstract}

\end{center}

In this work we employ the MHV technique to show that scattering amplitudes with any number of consecutive soft particles behave universally in the multi-soft limit in which all particles go
soft simultaneously.
After identifying the diagrams which give the leading contribution we give the general rules for writing down compact expressions for the multi-soft
factor of $m$ gluons, $k$ of which have
negative helicity. We explicitly consider the cases where $k=1$ and 2.
In $\cN =4$ SYM, the multi-soft factors of 2 scalars or 2 fermions forming a singlet of $SU(4)$ $R$-symmetry,
and $m-2$ positive helicity gluons are derived.
The special case of the double-soft limit gives an amplitude whose
leading divergence is $1/\delta^2$ and not $1/\delta$ as in the case of 2 scalars or 2 fermions that do not form a singlet under $SU(4)$.
The construction based on the analytic supervertices allows us to obtain simple expressions for the triple-soft limit of 1 scalar and 2 positive helicity fermions, as well as
for the quadrapole-soft limit of 4 positive helicity fermions, in a singlet configuration.

\vspace{0.5cm}

\setcounter{page}{0}
\thispagestyle{empty}
\newpage

\baselineskip 6mm

\section{Introduction}
On-shell scattering amplitudes is one of the most important observables calculated in quantum field theory\footnote{Another important class of observables
is that of correlation functions of gauge invariant operators. For some recent progress on this front see \cite{3-point} and references therein.}.
The complete knowledge of the scattering matrix
specifies the theory, at least at the perturbative level. Moreover, scattering amplitudes of gauge field
theories frequently exhibit structures and symmetries which are not at all apparent from
the conventional Lagrangian formulation of the theory\footnote{For detailed reviews on this and other related subjects see \cite{rev-amp}.}.

In particular, the soft behaviour of the S matrix of a theory generically reflects symmetries of this theory. One of the most important examples of this fact
is Weinberg's soft theorem \cite{soft-theor} which constrains the universal behaviour of the leading divergence
of the S-matrix in the soft limit by use of Ward identities. The universal behaviour of the subleading
terms in the momentum of the soft particle for gluons and gravitons have been also established in \cite{sublead}.
Another example, is that of the soft behaviour of Goldstone bosons
in the case of a spontaneous broken symmetry.
Taking the momentum of the Goldstone boson to approach zero leads to vanishing amplitude, a fact known as Adler's zero \cite{Adler}.
If one takes the double-soft limit of 2 scalars the resulting amplitude is non vanishing and can be related to some underlying symmetry of the theory.
For example, in the case of $\cN=8$ supergravity the double-soft limit reveals the $E_{7(7)}$ symmetry algebra of the scalar moduli space \cite{Kaplan}.
Similar double-soft theorems for spin $1/2$ particles and the restrictions that these may impose on supergravity counter terms have been also discussed in \cite{Chen:2014xoa}.
Furthermore, recently the subleading soft graviton theorems \cite{sublead} was argued to be related to large gauge transformations (diffeomorphisms in the case of gravity)
\cite{Strominger-grav}. The relevant asymptotic symmetry group should be the $BMS_4$ algebra \cite{BMS} or some extension of it in other than 4 space-time dimensions \cite{Palatrussardi,Strominger-grav,otherdim}.
The effect of loop corrections to the soft theorems was also considered in  \cite{soft-loops}.
More recently, the double-soft theorems for scalars and fermions have been extended to include the double-soft limits of gluons and gravitons \cite{Klose:2015xoa,Volovich:2015yoa}.
Further studies on the soft limit of the S matrix in gauge theories, gravity and string theory can be found in \cite{rest}.

In almost all the studies mentioned above the soft theorems were established through the BCFW recursion relations   \cite{BCFW} in 4 dimensions or through the CHY formula
in the case of arbitrary dimensions \cite{CHY}. In this work we will follow a slightly different route and employ the CSW (or scalar graph) method \cite{CSW}.
This method was originally proposed as an alternative to Feynman diagrammatics and was inspired by the weak-to-weak duality between $\cN =4$ SYM theory and the  topological
B model with target space the Calabi-Yau supermanifold $CP^{3|4}$ \cite{Witten}. This method was firstly proven to hold for tree-level gluonic amplitudes. Subsequently, it was proven to hold for all tree-level amplitudes in gauge theories with any amount of supersymmetry and for any number of colours \cite{GK,GGK}. The validity of the method at the loop level was also shown for the maximally supersymmetric theory, as well as in theories with
less supersymmetry \cite{MHV-loops,CSW2} \footnote{The equivalence of the MHV amplitudes with the expectation value of polygonal Wilson loops at both weak and strong coupling was also the subject of intense investigations \cite{me-wilson}.}.
The advantage of this approach is that it will allow us to easily prove multi-soft theorems involving an arbitrary number of soft particles.
Furthermore, it will also allow us to obtain compact expressions for the leading singularity of amplitudes involving any number of soft gluons, fermions and/or scalars.

The plan of the paper is the following. In Section 2 we discuss how the CSW method can be applied to the simultaneous multi-soft limit of an arbitrary number of gluons.
To be more precise the multi-soft limit that we will be considering is the following:
\be \label{soft-lim}
p_i=\delta q_i, \,\,\, \lambda_{p_i}^{\alpha}= \sqrt{\delta}\lambda_{q_i}^{\alpha}\equiv\sqrt{\delta}\lambda_{i}^{\alpha}, \,\,\,
{\tilde \lambda}_{p_i}^{\dot \alpha}= \sqrt{\delta}{\tilde \lambda}_{q_i}^{\dot \alpha}\equiv\sqrt{\delta}{\tilde \lambda}_{i}^{\dot \alpha},
\,\,\,\delta\rightarrow 0, \,\,\,i=1,...,m.
\ee
That is we take the momenta of m consecutive gluons to scale to zero simultaneously.
In particular, in Section 2.1 we prove  the multi-soft theorem and derive the multi-soft factor of m soft gluons one of which is of negative helicity. In Section 2.2 we give
the general rules from which one can obtain the multi-soft
factor of $m$ soft gluons $k$ of which have
negative helicity.
As an example, in Section 2.3 we apply these rules to write down the expression for the case of 2 negative and $m-2$ positive helicity gluons.
In all cases the result takes the form of the multi-soft limit of m positive helicity gluons times a factor depending on the negative helicity ones.
In Section 3 we consider the multi-soft limit in $\cN =4$ SYM theory.
After reviewing the scalar graph approach based on the analytic supervertices of the theory,
we proceed to consider the multi-soft limit of 2 scalars or 2 fermions forming a singlet under the $SU(4)$ R-symmetry group
and $m-2$ positive helicity gluons.
In Sections 3.3 and 3.4 we focus on the double soft limit of 2 scalars and 2 fermions respectively. We obtain the soft factor whose
leading divergence is $1/\delta^2$ and not $1/\delta$ as in the case of 2 scalars or 2 fermions that do not form a singlet under $SU(4)$.
In Section 3.5, we derive the expressions for the multi-soft function of (2 fermions + 1 scalar) and (4 fermions)
forming a singlet. In all cases the amplitude has a leading divergence of $1/\delta^m$ in the soft limit \eqref{soft-lim}.
 Finally, in Section 4 we conclude.

Before closing this Section a couple of comments are in order.
The objects that we will be dealing with will be the kinematic parts of
the colour-stripped scattering amplitudes. Furthermore, we will be suppressing the overall momentum conservation factor of the amplitudes.
Finally, the spinor conventions the we will adopt
are those of \cite{GK}.

\section{Gluonic multi-soft limits and the CSW method}

One way to obtain compact expressions for purely gluonic amplitudes with any number of particles
at tree level is by employing the CSW formalism \cite{CSW}.
In this Section we show how one can use this technique to get compact expressions for the multi-soft functions multiplying the "hard" amplitude
when any number of adjacent gluons are taken soft simultaneously, namely
\be \label{soft-lim}
p_i=\delta q_i, \,\,\, \lambda_{p_i}^{\alpha}= \sqrt{\delta}\lambda_{q_i}^{\alpha}\equiv\sqrt{\delta}\lambda_{i}^{\alpha}, \,\,\,
{\tilde \lambda}_{p_i}^{\dot \alpha}= \sqrt{\delta}{\tilde \lambda}_{q_i}^{\dot \alpha}\equiv\sqrt{\delta}{\tilde \lambda}_{i}^{\dot \alpha},
\,\,\,\delta\rightarrow 0, \,\,\,i=1,...,m.
\ee
When all these gluons are of the same helicity one can show that the multi-soft factor is the product of m
single soft factors as if the particles were taken soft one by one . This is no longer true when the multi-soft limit involves gluons
of different helicities \cite{Klose:2015xoa,Volovich:2015yoa}.
In this note we will restrict ourselves to the leading term in the $1/\delta$ expansion of the universal multi-soft factor $S_m$.
The first step consists in identifying the set of MHV diagrams which contribute in the soft limit. As we will see
only a limited number of MHV diagrams contribute in the soft limit \eqref{soft-lim}.

\subsection{Gluonic multi-soft factor with one negative helicity gluon}
\begin{figure}[h]
\label{fig1}
\begin{center}
{\scalebox{0.99}{
\includegraphics{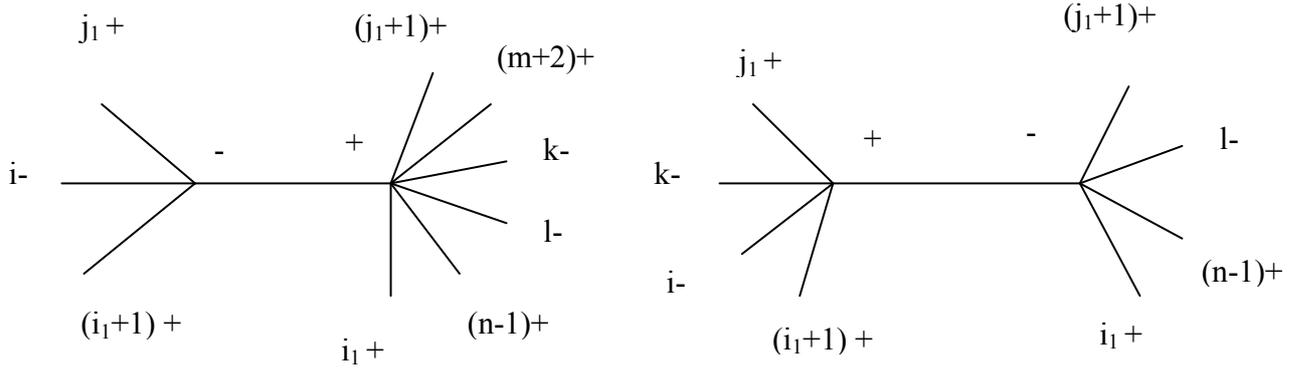}}
}
\end{center}
\caption{\it On the left tree diagrams with MHV vertices contributing to the leading
multi-soft factor
$S_m^{(0)}(1^+,2^+,...,i^-,...m^+)$.
On the right tree diagrams which give subleading contributions to the multi-soft factor. In the case of the factorisation of a generic "hard" amplitude the
right vertex should be "dressed" with additional MHV vertices. Apparently, this will not change the soft factor \eqref{1soft}, \eqref{12-fin}.  }
\end{figure}
In order to be able to extract the soft limit factor $S_m(1^+,2^+,...i^-,...m^+)$
we consider the NMHV n-gluon amplitude $A_n(1^+,...,i^-,...k^-,...,l^-,...,n)$.
We would like to stress that the reason for choosing the NMHV n-gluon amplitude is purely for simplicity.
As will become obvious from the diagrammatic analysis below, the same universal factorisation \eqref{1minus} is valid
for any chosen initial amplitude $A_n$. The only difference will be that the $A_{n-m}$ amplitude that factorises
will have more than 2 negative helicity gluons.
From now on we will omit writing explicitly the positive helicity gluons unless needed.
As discussed above in this Section we will take the simultaneous soft limit of m gluons of which only one will be
of negative helicity. Then the multi-soft factor can be identified from the following relation
\begin{eqnarray}\label{1minus}
&& \lim_{\delta \rightarrow 0}A_n(i^-,k^-,l^-)= S_m(1^+,2^+,...i^-,...m^+)A_{n-m}(m+1,...,k^-,...l^-,...,n)+{\mathcal O}({\delta}), \nonumber \\
&& S_m(1^+,2^+,...i^-,...m^+)=\sum_{a=0}^{m}\frac{1}{\delta^{m-a}}S_m^{(a)}(i^-).
\end{eqnarray}
In what follows we will concentrate on the leading term $S_m^{(0)}$ in the expansion of \eqref{1minus}. We should mention that
there is no obstacle in going beyond the leading order.

Actually, there are two classes of diagrams that can potentially contribute to $S_m^{(0)}$. They are depicted in Figure 1.
To start with notice that all the diagrams on the right of Figure 1, which have only one negative helicity gluon, either $k$ or $l$ on the right vertex are absent.
To convince ourselves lets us count the powers of $\delta$ of such a diagram.
The right MHV vertex will behave like $1/\delta^{i_1}$. This scaling comes from the fact the right MHV vertex has
$2 i_1$ holomorphic spinors in its denominator each of which scales as $\sqrt{\delta}$. Notice that when two or more of the
hard gluons appear in the left vertex then the propagator is finite in the multi-soft limit \eqref{soft-lim} and as a consequence it does not
contribute any power of $\delta$ in the leading term $S_m^{(0)}$ of the soft factor.
Similarly, the left MHV vertex behaves like $\delta^2/\delta^{m-i_1}$, where the additional $\delta^2$ in the numerator
is coming from the
$\vev{i~k}^4$ term present
in the numerator of the left MHV vertex. As a result, the whole diagram scales like $1/\delta^{m-2}$ which gives contribution only
in the sub-subleading soft factor $S_m^{(2)}$.
Actually we will see that this is a generic feature, namely all diagrams with more than two hard particles belonging to the same MHV vertex which has a negative helicity soft gluon
will not contribute to the leading and sub-leading soft factors $S_m^{(0)}$ and $S_m^{(1)}$ and can safely be ignored.

Using the expressions for the gluonic MHV vertices,
one can write down an analytic expression
for the
left diagram of Figure 1:
\SP{ \label{1neg}
A_n^{(1)} =
\sum_{i_1={n-1}}^{i-1\,\,\,\,\,'} \sum_{j_1=i}^{m+1\,\,\,\,\,'}{\vev{(i_1+1,j_1)~ {i}}^4 \over \vev{(i_1+1,j_1)~ i_1+1}...\vev{j_1-1~j_1}\vev{j_1~(i_1+1,j_1)}}
{1 \over  P_{i_1+1,j_1}^2} \nn \\
\times
{\vev{k~ {l}}^4 \over \vev{(i_1+1,j_1)~ j_1+1}...\vev{i_1-1~i_1}\vev{i_1~(i_1+1,j_1)}},}
where the primes at the sums are to remind us that we should omit the term for $i_1=n-1$ and $j_1=m+1$ since this term has
two hard gluons on the left MHV vertex and as argued above is subleading in the soft limit.
In the last expression the quantity $|(i_1+1,j_1)\rangle$ denotes the spinor that corresponds to the off-shell momentum of the propagator $P_{i_1+1,j_1}$.
It is defined as usual $\lambda _{(i_1+1,j_1)\,\,\,\alpha}=P_{i_1+1,j_1\,\,\,\alpha \dot\alpha}  \zeta^{\dot\alpha}$, where $\zeta^{\dot\alpha}$ is an arbitrary
reference spinor.

Taking the multi-soft limit \eqref{soft-lim} this expression can be easily rewritten as
\SP{ \label{1negpre}
A_n^{(1)} ={ \vev{n~m+1}\over \prod_{q=n}^{m} \vev{q~q+1}}
\sum_{i_1={n-1}}^{i-1\,\,\,\,\,'} \sum_{j_1=i}^{m+1\,\,\,\,\,'}{\vev{i^-|{p\!\!\!/}_{i_1+1,j_1}| \zeta^-}^4 \over D}A_{n-m}(m+1^+,...k^-,l^-,...,n^+).}
where we have defined the universal combination,
\EQ{ \label{Ddef}
D
=\vev{i_1^-|{p\!\!\!/}_{i_1+1,j_1}| \zeta^-}
\vev{j_1+1^-|{p\!\!\!/}_{i_1+1,j_1}| \zeta^-}
\vev{i_1+1^-|{p\!\!\!/}_{i_1+1,j_1}| \zeta^-}
\vev{j_1^-|{p\!\!\!/}_{i_1+1,j_1}| \zeta^-}\,
\frac{P_{i_1+1,j_1}^2}{\vev{i_1~i_1+1}\vev{j_1~j_1+1}},
}
and, as usual, $\zeta$ is the reference spinor of the CSW method.

From \eqref{1negpre} it is direct to read the multi-soft factor that we are after.
It reads
\SP{ \label{1soft}
S_m^{(0)}(i^-) ={ \vev{n~m+1}\over \prod_{q=n}^{m} \vev{q~q+1}}
\sum_{i_1={n-1}}^{ i-1\,\,\,\,\,'} \sum_{j_1=i}^{ m+1\,\,\,\,\,'}{\vev{i^-|{p\!\!\!/}_{i_1+1,j_1}| \zeta^-}^4 \over D}.}
A number of important comments are in order.
First of all, notice that our result \eqref{1soft} appears to depend on the arbitrary reference spinor $\zeta$.
We will now argue that the result \eqref{1soft} is actually independent of the reference spinor $\zeta$.
On one hand, we know that the full amplitude is independent of the value of the reference spinor \cite{CSW}.
As a result, any multi-soft limiting process should give a result that should be also independent of the reference spinor.
In fact, each term $S_m^{(a)}(i^-)$ in the $1/\delta$ expansion of $S_m(i^-)$ \eqref{1minus} should be separately independent
since the cancellations are only possible between terms that have the same power divergence $1/\delta^a$ in the multi-soft limit.
We conclude that the soft factor we have calculated and which is the leading singularity of the full amplitude should not depend
on which reference spinor one chooses. Below we will verify this for the simple case of the double-soft limit.
In practice, we will choose $\zeta$ to be the anti-holomorphic spinor of a positive helicity gluon, namely
$\zeta^{\dot \a}={\tilde \lambda}_n^{\dot \a}$ or $\zeta^{\dot \a}={\tilde \lambda}_1^{\dot \a}$. As a consequence, our result
will be Lorentz and gauge invariant and free of singularities connected to the choice of the reference spinor.
A second comment concerns the structure of our result \eqref{1soft}. Notice that this result is the expression
for the multi-soft limit of $m$ adjacent positive helicity gluons multiplied by a correction term given by the double sum.

The careful reader might worry that the expression \eqref{1soft} depends not only on the momenta of the soft gluons $g_1,...,g_m$
and their adjacent particles $n$ and $m+1$, as it should, but also on the momenta of the next-to-adjacent gluons, i.e. $n-1$ and $m+2$
since the sums \eqref{1soft} involve these gluons. To clarify this point, one should evaluate the boundary terms of \eqref{1minus},
that is the terms for $i_1=n-1$ or $j_1=m+1$. By doing so it is easy to see that the dependence on the $n-1$ and $m+2$ gluons drops out.
Indeed, for $i_1=n-1$ the double sum of \eqref{1soft} simplifies to
\SP{ \label{1softb1}
S_m^{(i_1=n-1)}(i^-) ={ \vev{n~m+1}\over \prod_{q=n}^{m} \vev{q~q+1}}
\sum_{j_1=i}^{ m}{\vev{i^-|{p\!\!\!/}_{n}| \zeta^-}^4 \over D},}
with the denominator being
\EQ{ \label{n-1}
D
=[n \zeta]
\vev{j_1+1^-|{p\!\!\!/}_{n}| \zeta^-}
\vev{n^-|{q\!\!\!/}_{1,j_1}| \zeta^-}
\vev{j_1^-|{p\!\!\!/}_{n}| \zeta^-}\,
\frac{2 p_n \cdot q_{1,j_1}}{\vev{j_1~j_1+1}}.
}
Thus the last two expressions are explicitly independent of the gluon $n-1$.\\
Similarly, for the other boundary term $j_1=m+1$ we get
\SP{ \label{1softb1}
S_m^{(j_1=m+1)}(i^-) ={ \vev{n~m+1}\over \prod_{q=n}^{m} \vev{q~q+1}}
\sum_{i_1=n}^{ i-1}{\vev{i^-|{p\!\!\!/}_{m+1}| \zeta^-}^4 \over D},}
with the denominator being
\EQ{ \label{m+1}
D
=-\vev{i_1^-|{p\!\!\!/}_{m+1}| \zeta^-}
[m+1 \zeta]
\vev{i_1+1^-|{p\!\!\!/}_{m+1}| \zeta^-}
\vev{m+1^-|{p\!\!\!/}_{i_1+1,m}| \zeta^-}\,
\frac{2 p_{m+1}\cdot q_{i_1+1,m}}{\vev{i_1~i_1+1}},
}
which gives a result independent of the  details of the $m+2$ gluon.

One can use the general expression \eqref{1soft} to obtain the double soft factor when the gluons $1^+$ and $2^-$ become soft simultaneously.
There are four terms which after a bit of algebra read
\bea
 \label{12-}
S_2^{(a)}(1^+,2^-) ={ \vev{n~3}\over  \vev{1~2} q_{12}^2}
{\vev{2^-|{p\!\!\!/}_{1}| \zeta^-}^3 \over \vev{1^-|{p\!\!\!/}_{2}| \zeta^-}\vev{3^-|{q\!\!\!/}_{12}| \zeta^-}\vev{n^-|{q\!\!\!/}_{12}| \zeta^-}}  \nn \\
S_2^{(b)}(1^+,2^-) =-{ \vev{n~3}\over  \vev{1~2}\vev{2~3} 2 q_{12}\cdot p_3}
{\vev{2^-|{p\!\!\!/}_{3}| \zeta^-}^4 \over \vev{1^-|{p\!\!\!/}_{3}| \zeta^-} \vev{3^-|{q\!\!\!/}_{12}| \zeta^-}\vev{n^-|{p\!\!\!/}_{3}| \zeta^-}[3\zeta]} \nn \\
S_2^{(c)}(1^+,2^-) =-{ \vev{n~3}\over  \vev{n~1}\vev{2~3} 2 q_{2}\cdot p_3}
{\vev{2^-|{p\!\!\!/}_{3}| \zeta^-}^3 \over \vev{3^-|{p\!\!\!/}_{2}| \zeta^-} \vev{1^-|{p\!\!\!/}_{3}| \zeta^-}[3\zeta]} \nn \\
S_2^{(d)}(1^+,2^-) ={ \vev{n~3}\over  \vev{1~2}\vev{n~1} 2 q_{12}\cdot p_n}
{\vev{2^-|{p\!\!\!/}_{n}| \zeta^-}^3 \over \vev{3^-|{p\!\!\!/}_{n}| \zeta^-} \vev{n^-|{q\!\!\!/}_{12}| \zeta^-}[n\zeta]}
\eea
We have checked numerically that the sum of the four terms above is indeed independent of the reference spinor $\zeta^{\dot \a}$.

It is convenient to choose as the reference spinor that of first positive helicity gluon $1^+$, that is $\zeta^{\dot \a}={\tilde \lambda}_1^{\dot \a}$.
Then the first contribution $S_2^{(a)}(1^+,2^-)=0$ vanishes and the final result is a sum of the last three terms in \eqref{12-} after the substitution $\zeta^{\dot \a}={\tilde \lambda}_1^{\dot \a}$.
After some simple algebra we get the final result
\bea
 \label{12-fin}
S_2(1^+,2^-) ={ 1\over  \vev{1~2} [21]} \Big({\vev{2~3}^2 [31]\over \vev{1~3} 2 q_{12}\cdot p_3}+{\vev{1~2}\vev{2~3}\vev{n~3}[31]\over \vev{n~1} \vev{1~3} 2 q_{2}\cdot p_3}+{\vev{2~n}^2 [n1]\over \vev{n~1}
2 q_{12}\cdot p_n}\Big).
\eea
One can check that this result is another representation of the double soft limit
\bea
 \label{Gab}
 S_2(1^+,2^-)=\frac{1}{\vev{n^-|{q\!\!\!/}_{12}| 3^-}}(\frac{1}{2 p_n\cdot q_{12}} { \vev{n~2}^3 [n3]\over  \vev{n~1}\vev{1~2}}-
 \frac{1}{2 p_3\cdot q_{12}} { \vev{n~3} [31]^3\over [12][23]})
 \eea
first obtained in  \cite{Volovich:2015yoa,Klose:2015xoa}.

At this point we wish to comment on the effectiveness of the CSW method compared to this of the BCFW method.
First of all from the CSW method we directly get final expression for the multi-soft factor and not recursive relations.
Secondly, if we roughly estimate the number of terms in the sums of the CSW method this will be the number of diagrams with different topology times
$m^{2n}$, where $m$ is the number of soft particles and $n$ the number of negative helicity gluons since each negative helicity gluon brings in a double sum (see \eqref{1soft} and the results of the next Sections). On the other hand, due to the recursive nature of the BCFW method in order to calculate the $m^{th}$ multi-soft limit one has a sum of $m-1$ terms corresponding to all multi-soft
factors from $S_2$ to $S_{m-1}$. Now each of these multi-soft
factors is written as a sum of all the lower point ones and so on. As a result, the number of terms in the final result of the BCFW method grows exponentially ($\sim 2^m$) with respect to the number of soft particles. Thus for $m>>1$ and $n<<m/2$ the number of terms in the CSW result is much smaller than the one in the BCFW method.

\subsection{Rules for generic gluonic multi-soft factor }

In this Section, we show how to construct a generic multi-soft factor $S_m((i_1)^-,(i_2)^-,...,(i_k)^-)$ with the aid of MHV diagrams.
The limit we take is the simultaneous soft limit of $m$ gluons $k$ of which are of negative helicity.
Having the experience of the simple example in the previous Section it is not difficult
to write down the general rules for such a multi-soft limit.
We should stress that these rules are independent of which "hard" amplitude factorises and multiplies the multi-soft factor
$S_m$ when the multi-soft limit is taken.
As a result the multi-soft factorisation is universal with the amplitude that factorises being independent of the momenta of the soft gluons
and with the multi-soft factor $S_m$ depending only on the momenta of the soft particles,
as well as on the two hard gluons which are adjacent to the soft ones,
i.e. the $n$ and $m+1$ gluons. \\
\\
1. Draw all different topologically different "skeleton" MHV diagrams each with $k-1$ MHV vertices.  \\
\\
2. Distribute the negative helicity gluons among the MHV vertices of each "skeleton" diagram in all inequivalent ways.
Then distribute all positive helicity gluons in all possible ways respecting the ordering of the particles. The particles which one needs to
distribute are the soft ones plus the four hard gluons adjacent to the m soft ones, that is $n,n-1,m+1,m+2$.
In accordance with the discussion
in the previous Section, care should be taken
so that the gluons $m+2$ and $n-1$ belong to the same MHV vertex. Otherwise the diagram will give only subleading contribution to the multi-soft factor
\footnote{ As discussed in the previous Section the dependence on the gluons $m+2$ and $n-1$ drops out of the final result.} .\\
\\
3. For each MHV vertex, except the one which has the $n-1$ and $m+2$ gluons,  include a factor of $\vev{n_1~n_2}^4$ in the numerator, where $n_1$ and $n_2$ are the negative helicity gluons (internal or external) of the MHV vertex. \\
\\
4. For each propagator of momentum $P_i$ connecting two MHV vertices include a factor $1/D_i$, where
\EQ{ \label{Di}
D_i
=\vev{j_1^-|{P\!\!\!/}_{i}| \zeta^-}
\vev{j_3^-|{P\!\!\!/}_{i}| \zeta^-}
\vev{j_4^-|{P\!\!\!/}_{i}| \zeta^-}
\vev{j_2^-|{P\!\!\!/}_{i}| \zeta^-}\,
P_{i}^2,}
where $j_1,j_2,j_3,j_4$ are the four gluons adjacent to the propagator under consideration.
Notice that it might happen that one or more
of these gluons are propagators too. If a diagram has propagators which are adjacent multiply by
$\vev{P_i~P_j}=-\vev{\zeta^+|{P\!\!\!/}_{i}{P\!\!\!/}_{j}| \zeta^-}$ for each
set of adjacent propagators $(P_i,P_j)$. \\
\\
5. Multiply by a factor $\vev{j_1~j_1+1}$ for all adjacent external particles which do not belong to the same MHV vertex.\\
\\
6. Finally, sum over the contributions of all diagrams and multiply by the ubiquitous factor $U={ \vev{n~m+1}\over \prod_{q=n}^{m} \vev{q~q+1}}$.

\subsection{Gluonic multi-soft factor with two negative helicity gluons}

\begin{figure}[h]
\label{fig2}
\begin{center}
{\scalebox{0.85}{
\includegraphics{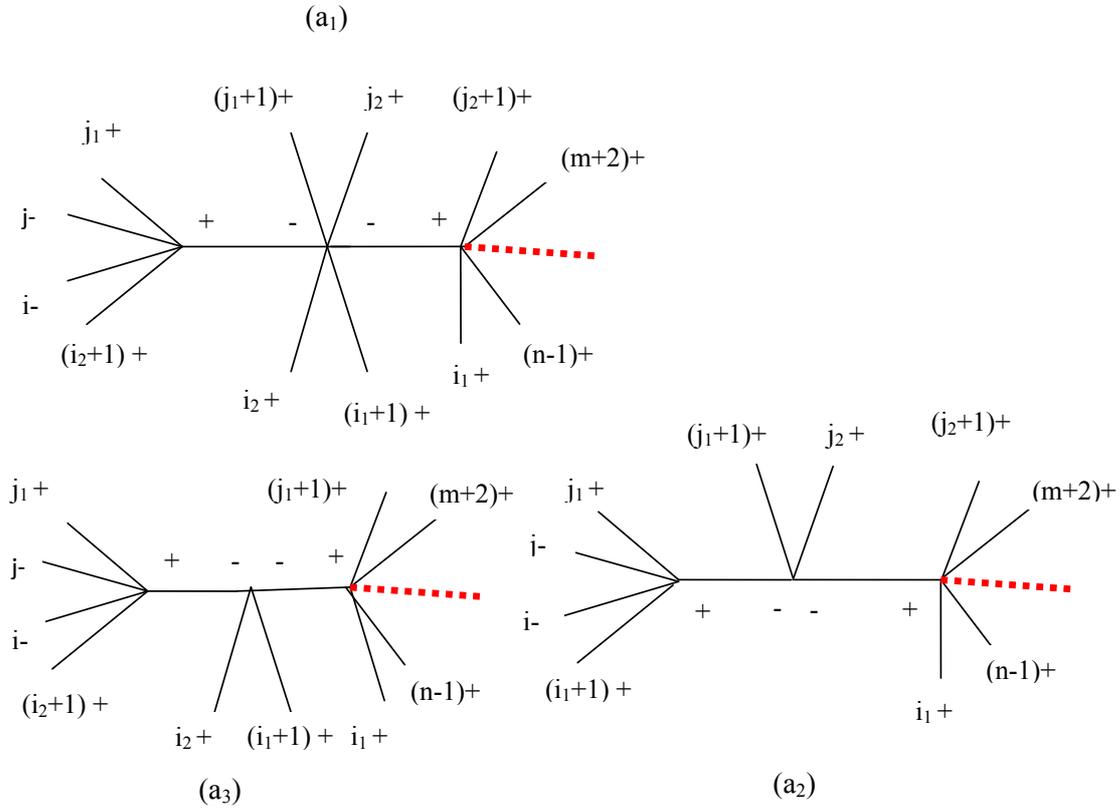}}
}
\end{center}
\caption{\it Diagrams contributing to the multi-soft factor involving 2 negative helicity gluons $S_m^{(0)}(1^+,...,i^-,...,j^-,...,m^+)$. The negative helicity gluons are on the left MHV vertex. The red dashed line denotes the additional gluons and possibly vertices of the hard amplitude that factorises in the soft limit.
}
\end{figure}
In this Section, as an example of the aforementioned rules we give compact expressions for the leading singularity of the
gluonic multi-soft factor with two negative and $m-2$ positive helicity gluons. In this Section, our choice for the reference spinor will be $\zeta^{\dot \alpha}=\lambda_n^{\dot \alpha}$.

There is a total of nine topologically different diagrams which contribute to this multi-soft limit. They are depicted in Figures 2, 3 and 4.
For the fist diagram (a1) of Figure 2 we get
\SP{ \label{a1}
S_m^{(a_1)}(i^-,j^-) =U
\sum_{i_1={n-1}}^{ i-2\,\,\,\,\,'} \sum_{i_2={i_1+1}}^{ i-1\,\,\,\,\,'}
\sum_{j_1=j}^{ m\,\,\,\,\,'} \sum_{j_2={j_1+1}}^{ m+1\,\,\,\,\,'}
{\vev{i~j}^4\vev{n^+|{p\!\!\!/}_{i_1+1,j_2}{p\!\!\!/}_{i_2+1,j_1}| n^-}^4 \over D_1 D_2}\times \\ \vev{i_1~i_1+1}\vev{i_2~i_2+1}\vev{j_1~j_1+1}
 \vev{j_2~j_2+1},}
where
\SP{
D_1=\vev{i_2^-|{P\!\!\!/}_{i_2+1,j_1}| n^-}
\vev{i_2+1^-|{P\!\!\!/}_{i_2+1,j_1}| n^-}
\vev{j_1^-|{P\!\!\!/}_{i_2+1,j_1}| n^-}
\vev{j_1+1^-|{P\!\!\!/}_{i_2+1,j_1}| n^-}\,
P_{i_2+1,j_1}^2,\,\,\,\,\\
D_2=\vev{i_1^-|{P\!\!\!/}_{i_1+1,j_2}| n^-}
\vev{i_1+1^-|{P\!\!\!/}_{i_1+1,j_2}| n^-}
\vev{j_2+1^-|{P\!\!\!/}_{i_1+1,j_2}| n^-}
\vev{j_2^-|{P\!\!\!/}_{i_1+1,j_2}| n^-}\,
P_{i_1+1,j_2}^2
}

The other two diagrams (a2) and (a3) of Figure 2 yield
\SP{ \label{a2}
S_m^{(a_2)}(i^-,j^-) =U
\sum_{i_1={n-1}}^{ i-1\,\,\,\,\,'}
\sum_{j_1=j}^{ m\,\,\,\,\,'} \sum_{j_2={j_1+1}}^{ m+1\,\,\,\,\,'}
{\vev{i~j}^4\vev{n^+|{P\!\!\!/}_{i_1+1,j_2}{P\!\!\!/}_{i_1+1,j_1}| n^-}^3 \over D_1 D_2}\times \\
\vev{i_1~i_1+1}\vev{j_1~j_1+1} \vev{j_2~j_2+1},}
where
\SP{
D_1=
\vev{i_1+1^-|{P\!\!\!/}_{i_1+1,j_1}| n^-}
\vev{j_1^-|{P\!\!\!/}_{i_1+1,j_1}| n^-}
\vev{j_1+1^-|{P\!\!\!/}_{i_1+1,j_1}| n^-}\,
P_{i_1+1,j_1}^2,\,\,\,\,\\
D_2=\vev{i_1^-|{P\!\!\!/}_{i_1+1,j_2}| n^-}
\vev{j_2+1^-|{P\!\!\!/}_{i_1+1,j_2}| n^-}
\vev{j_2^-|{P\!\!\!/}_{i_1+1,j_2}| n^-}\,
P_{i_1+1,j_2}^2
}

\SP{ \label{a3}
S_m^{(a_3)}(i^-,j^-) =-U
\sum_{i_1={n-1}}^{ i-2\,\,\,\,\,'}\sum_{i_2={i_1+1}}^{ i-1\,\,\,\,\,'}
\sum_{j_1=j}^{ m+1\,\,\,\,\,'}
{\vev{i~j}^4\vev{n^+|{P\!\!\!/}_{i_1+1,j_1}{P\!\!\!/}_{i_2+1,j_1}| n^-}^3 \over D_1 D_2}\times \\
\vev{i_1~i_1+1}\vev{j_1~j_1+1} \vev{i_2~i_2+1},}
where
\SP{
D_1=
\vev{i_2^-|{P\!\!\!/}_{i_2+1,j_1}| n^-}
\vev{j_1^-|{P\!\!\!/}_{i_2+1,j_1}| n^-}
\vev{i_2+1^-|{P\!\!\!/}_{i_2+1,j_1}| n^-}\,
P_{i_2+1,j_1}^2,\,\,\,\,\\
D_2=\vev{i_1^-|{P\!\!\!/}_{i_1+1,j_1}| n^-}
\vev{j_1+1^-|{P\!\!\!/}_{i_1+1,j_1}| n^-}
\vev{i_1+1^-|{P\!\!\!/}_{i_1+1,j_1}| n^-}\,
P_{i_1+1,j_1}^2
}

\begin{figure}[h]
\label{fig3}
\begin{center}
{\scalebox{0.83}{
\includegraphics{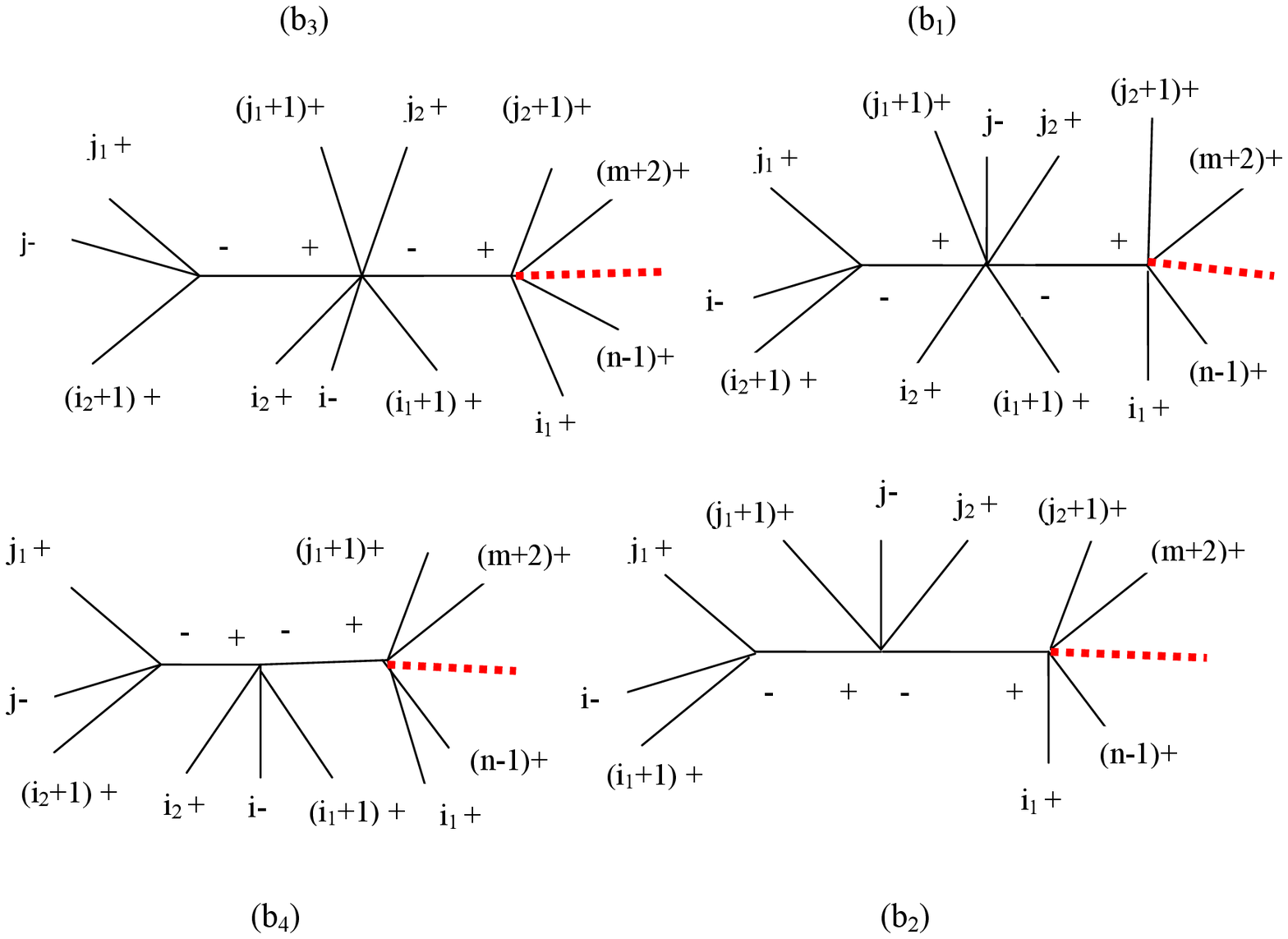}}
}
\end{center}
\caption{\it Diagrams contributing to the multi-soft factor involving 2 negative helicity gluons $S_m^{(0)}(1^+,...,i^-,...,j^-,...,m^+)$. Here the negative helicity gluons are sitting
one on the left MHV vertex and one on the middle MHV vertex. The red dashed line denotes the additional gluons and possibly vertices of the hard amplitude that factorises in the soft limit.
}
\end{figure}

We now proceed to evaluate the four diagrams of Figure 3.
These give
\SP{ \label{b1}
S_m^{(b_1)}(i^-,j^-) =U
\sum_{i_1={n-1}}^{ i-2\,\,\,\,\,'} \sum_{i_2={i_1+1}}^{ i-1\,\,\,\,\,'}
\sum_{j_1=i}^{ j-1\,\,\,\,\,'} \sum_{j_2=j}^{ m+1\,\,\,\,\,'}
{\vev{i^-|{P\!\!\!/}_{i_2+1,j_1}| n^-}^4\vev{j^-|{P\!\!\!/}_{i_1+1,j_2}| n^-}^4 \over D_1 D_2}\times \\ \vev{i_1~i_1+1}\vev{i_2~i_2+1}\vev{j_1~j_1+1}
 \vev{j_2~j_2+1},}
where
\SP{
D_1=\vev{i_2^-|{P\!\!\!/}_{i_2+1,j_1}| n^-}
\vev{i_2+1^-|{P\!\!\!/}_{i_2+1,j_1}| n^-}
\vev{j_1^-|{P\!\!\!/}_{i_2+1,j_1}| n^-}
\vev{j_1+1^-|{P\!\!\!/}_{i_2+1,j_1}| n^-}\,
P_{i_2+1,j_1}^2,\,\,\,\,\\
D_2=\vev{i_1^-|{P\!\!\!/}_{i_1+1,j_2}| n^-}
\vev{i_1+1^-|{P\!\!\!/}_{i_1+1,j_2}| n^-}
\vev{j_2+1^-|{P\!\!\!/}_{i_1+1,j_2}| n^-}
\vev{j_2^-|{P\!\!\!/}_{i_1+1,j_2}| n^-}\,
P_{i_1+1,j_2}^2
}

\SP{ \label{b2}
S_m^{(b_2)}(i^-,j^-) =U
\sum_{i_1={n-1}}^{ i-1\,\,\,\,\,'}
\sum_{j_1=i}^{ j-1\,\,\,\,\,'} \sum_{j_2=j}^{ m+1\,\,\,\,\,'}
{\vev{i^-|{P\!\!\!/}_{i_1+1,j_1}| n^-}^4\vev{j^-|{P\!\!\!/}_{i_1+1,j_2}| n^-}^4 \over D_1 D_2}\times \\ \vev{i_1~i_1+1}\vev{j_1~j_1+1}
 \vev{j_2~j_2+1},}
where
\SP{
D_1=\vev{i_1+1^-|{P\!\!\!/}_{i_1+1,j_1}| n^-}
\vev{j_1^-|{P\!\!\!/}_{i_1+1,j_1}| n^-}
\vev{j_1+1^-|{P\!\!\!/}_{i_1+1,j_1}| n^-}\,
P_{i_1+1,j_1}^2,\,\,\,\,\\
D_2=\vev{i_1^-|{P\!\!\!/}_{i_1+1,j_2}| n^-}
\vev{n^+|{P\!\!\!/}_{i_1+1,j_2}{P\!\!\!/}_{i_1+1,j_1}| n^-}
\vev{j_2+1^-|{P\!\!\!/}_{i_1+1,j_2}| n^-}
\vev{j_2^-|{P\!\!\!/}_{i_1+1,j_2}| n^-}\,
P_{i_1+1,j_2}^2
}

\SP{ \label{b3}
S_m^{(b_3)}(i^-,j^-) =U
\sum_{i_1={n-1}}^{ i-1\,\,\,\,\,'} \sum_{i_2=i}^{ j-1\,\,\,\,\,'}
\sum_{j_1=j}^{ m\,\,\,\,\,'} \sum_{j_2=j_1+1}^{ m+1\,\,\,\,\,'}
{\vev{j^-|{P\!\!\!/}_{i_2+1,j_1}| n^-}^4\vev{i^-|{P\!\!\!/}_{i_1+1,j_2}| n^-}^4 \over D_1 D_2}\times \\ \vev{i_1~i_1+1}\vev{i_2~i_2+1}\vev{j_1~j_1+1}
 \vev{j_2~j_2+1},}
where
\SP{
D_1=\vev{i_2^-|{P\!\!\!/}_{i_2+1,j_1}| n^-}
\vev{i_2+1^-|{P\!\!\!/}_{i_2+1,j_1}| n^-}
\vev{j_1^-|{P\!\!\!/}_{i_2+1,j_1}| n^-}
\vev{j_1+1^-|{P\!\!\!/}_{i_2+1,j_1}| n^-}\,
P_{i_2+1,j_1}^2,\,\,\,\,\\
D_2=\vev{i_1^-|{P\!\!\!/}_{i_1+1,j_2}| n^-}
\vev{i_1+1^-|{P\!\!\!/}_{i_1+1,j_2}| n^-}
\vev{j_2+1^-|{P\!\!\!/}_{i_1+1,j_2}| n^-}
\vev{j_2^-|{P\!\!\!/}_{i_1+1,j_2}| n^-}\,
P_{i_1+1,j_2}^2
}

\SP{ \label{b4}
S_m^{(b_4)}(i^-,j^-) =-U
\sum_{i_1={n-1}}^{ i-1\,\,\,\,\,'} \sum_{i_2=i}^{ j-1\,\,\,\,\,'}
\sum_{j_1=j}^{ m+1\,\,\,\,\,'}
{\vev{j^-|{P\!\!\!/}_{i_2+1,j_1}| n^-}^4\vev{i^-|{P\!\!\!/}_{i_1+1,j_1}| n^-}^4 \over D_1 D_2}\times \\ \vev{i_1~i_1+1}\vev{i_2~i_2+1}\vev{j_1~j_1+1}
,}
where
\SP{
D_1=\vev{i_2^-|{P\!\!\!/}_{i_2+1,j_1}| n^-}
\vev{i_2+1^-|{P\!\!\!/}_{i_2+1,j_1}| n^-}
\vev{j_1^-|{P\!\!\!/}_{i_2+1,j_1}| n^-}\,
P_{i_2+1,j_1}^2,\,\,\,\,\\
D_2=\vev{i_1^-|{P\!\!\!/}_{i_1+1,j_1}| n^-}
\vev{i_1+1^-|{P\!\!\!/}_{i_1+1,j_1}| n^-}
\vev{j_1+1^-|{P\!\!\!/}_{i_1+1,j_1}| n^-}
\vev{n^+|{P\!\!\!/}_{i_2+1,j_1}{P\!\!\!/}_{i_1+1,j_1}| n^-}\,
P_{i_1+1,j_1}^2
}
\begin{figure}[h]
\label{fig3}
\begin{center}
{\scalebox{0.89}{
\includegraphics{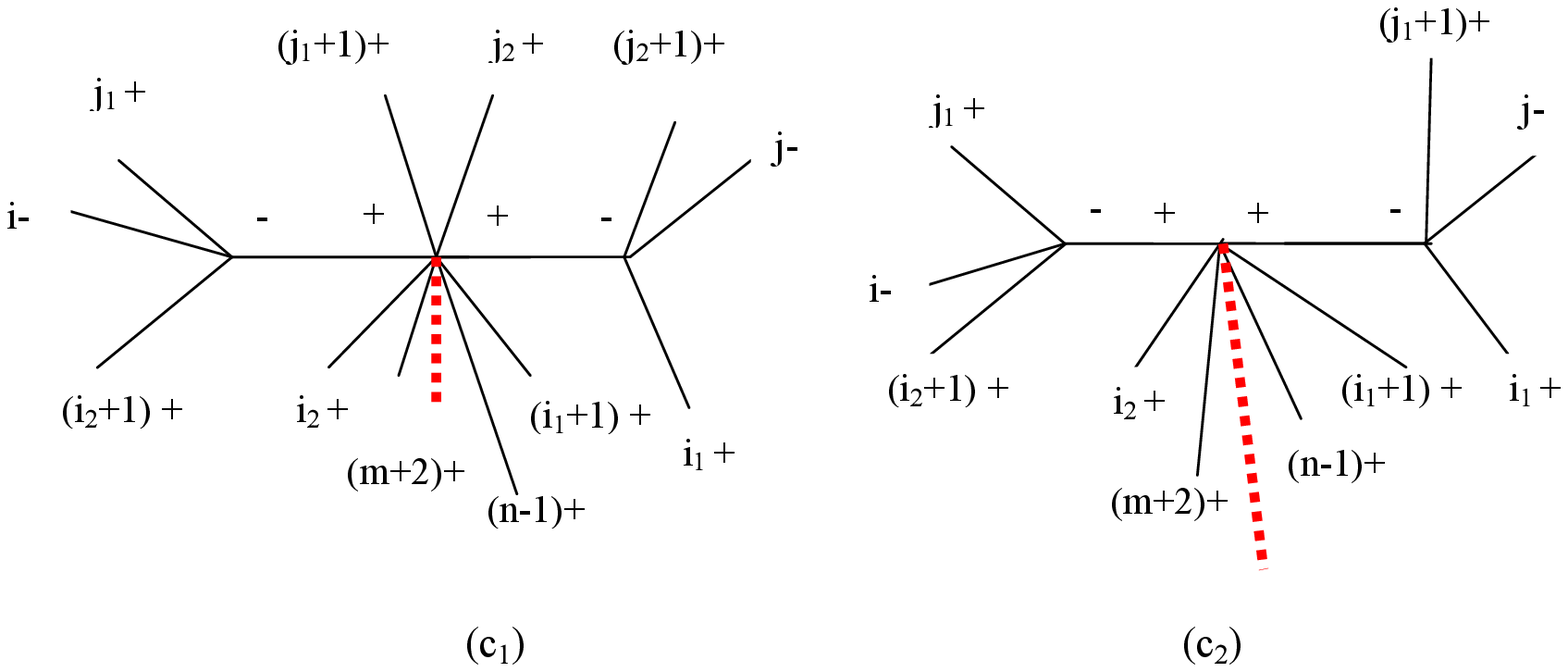}}
}
\end{center}
\caption{\it Diagrams contributing to the multi-soft factor involving 2 negative helicity gluons $S_m^{(0)}(1^+,...,i^-,...,j^-,...,m^+)$. Here the negative helicity gluons are sitting
one on the left MHV vertex and one on the right MHV vertex. The red dashed line denotes the additional gluons and possibly vertices of the hard amplitude that factorises in the soft limit.
}
\end{figure}

Finally, from the diagrams of Figure 3 one gets
\SP{ \label{c1}
S_m^{(c_1)}(i^-,j^-) =U
\sum_{i_1={j}}^{ n-1\,\,\,\,\,'} \sum_{i_2=m+2}^{ i-1\,\,\,\,\,'}
\sum_{j_1=i}^{ j-2\,\,\,\,\,'} \sum_{j_2=j_1+1}^{ j-1\,\,\,\,\,'}
{\vev{j^-|{P\!\!\!/}_{j_2+1,i_1}| n^-}^4\vev{i^-|{P\!\!\!/}_{i_2+1,j_1}| n^-}^4 \over D_1 D_2}\times \\ \vev{i_1~i_1+1}\vev{i_2~i_2+1}\vev{j_1~j_1+1}
 \vev{j_2~j_2+1},}
where
\SP{
D_1=\vev{i_2^-|{P\!\!\!/}_{i_2+1,j_1}| n^-}
\vev{i_2+1^-|{P\!\!\!/}_{i_2+1,j_1}| n^-}
\vev{j_1^-|{P\!\!\!/}_{i_2+1,j_1}| n^-}
\vev{j_1+1^-|{P\!\!\!/}_{i_2+1,j_1}| n^-}\,
P_{i_2+1,j_1}^2,\,\,\,\,\\
D_2=\vev{i_1^-|{P\!\!\!/}_{j_2+1,i_1}| n^-}
\vev{i_1+1^-|{P\!\!\!/}_{j_2+1,i_1}| n^-}
\vev{j_2+1^-|{P\!\!\!/}_{j_2+1,i_1}| n^-}
\vev{j_2^-|{P\!\!\!/}_{j_2+1,i_1}| n^-}\,
P_{j_2+1,i_1}^2
}

\SP{ \label{c2}
S_m^{(c_2)}(i^-,j^-) =-U
\sum_{i_1={j}}^{ n-1\,\,\,\,\,'} \sum_{i_2=m+2}^{ i-1\,\,\,\,\,'}
\sum_{j_1=i}^{ j-1\,\,\,\,\,'}
{\vev{j^-|{P\!\!\!/}_{j_1+1,i_1}| n^-}^4\vev{i^-|{P\!\!\!/}_{i_2+1,j_1}| n^-}^4 \over D_1 D_2}\times \\ \vev{i_1~i_1+1}\vev{i_2~i_2+1}\vev{j_1~j_1+1},}
where
\SP{
D_1=\vev{i_2^-|{P\!\!\!/}_{i_2+1,j_1}| n^-}
\vev{i_2+1^-|{P\!\!\!/}_{i_2+1,j_1}| n^-}
\vev{j_1^-|{P\!\!\!/}_{i_2+1,j_1}| n^-}\,
P_{i_2+1,j_1}^2,\,\,\,\,\\
D_2=\vev{i_1^-|{P\!\!\!/}_{j_1+1,i_1}| n^-}
\vev{i_1+1^-|{P\!\!\!/}_{j_1+1,i_1}| n^-}
\vev{j_1+1^-|{P\!\!\!/}_{j_1+1,i_1}| n^-}
\vev{n^+|{P\!\!\!/}_{j_1+1,i_1}{P\!\!\!/}_{i_2+1,j_1}| n^-}\,
P_{j_1+1,i_1}^2
}

Overall the final result for the multi-soft factor with two negative helicity gluons reads
$S_m(i^-,j^-)=S_m^{(a_1)}(i^-,j^-)+S_m^{(a_2)}(i^-,j^-)+S_m^{(a_3)}(i^-,j^-)+S_m^{(b_1)}(i^-,j^-)+S_m^{(b_2)}(i^-,j^-)
+S_m^{(b_3)}(i^-,j^-)+S_m^{(b_4)}(i^-,j^-)+S_m^{(c_1)}(i^-,j^-)+S_m^{(c_2)}(i^-,j^-)$.

\section{Multi-soft limits in $N=4$ SYM from the analytic supervertex}

\subsection{Iterating the analytic supervertex}
\begin{figure}[ht]
\label{fig6}
\psfrag{i+1}{\large$(j_1+1)$}
\psfrag{i}{\large$j_1$}
\psfrag{j+1}{\large$(i_1+1)$}
\psfrag{j}{\large$i_1$}
\psfrag{n1}{\large$n_1$}
\psfrag{n2}{\large$n_2$}
\psfrag{n+}{\large$n\,+$}
\psfrag{ib}{\large$\bar{I}$}
\psfrag{I}{\large$I$}
\psfrag{2-}{\large$2\,-$}
\psfrag{m2-}{\large$m_2\,-$}
\psfrag{m3-}{\large$m_3\,-$}
\psfrag{+}{\large$+$}
\psfrag{-}{\large$-$}
\begin{center}
{\scalebox{0.5}{
\includegraphics{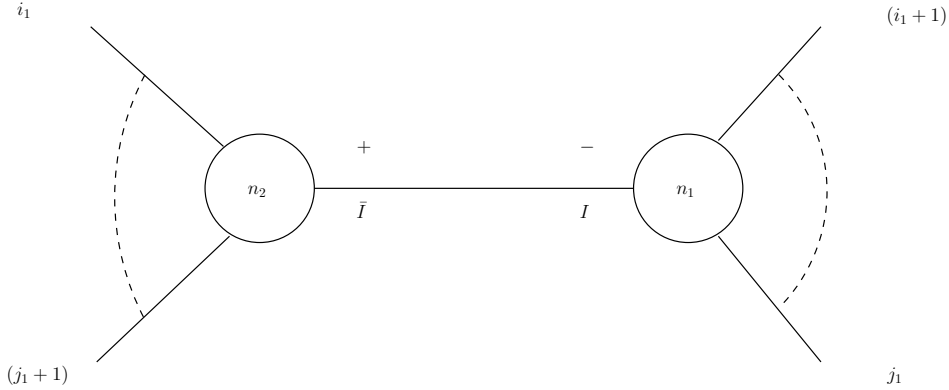}}
}
\end{center}
\caption{\it Tree diagrams with 2 analytic  supervertices contributing to the
degree-12 superamplitude of
Eq.~\eqref{finrest}. $n_1$ and $n_2$ are the number of legs in the right and left vertex respectively.
}
\end{figure}

In this Section, we briefly review the application of the CSW method in $\cN=4$ SYM
following closely \cite{GGK}. The building block of this construction will be the
analytic supervertex of Nair \cite{Nair}.

It is well-known that all analytic amplitudes in generic $0\le \cN \le 4$
gauge theories can be combined into a single $\cN=4$ supersymmetric
expression
given by \cite{Nair},
\be
A_n^{\cN=4} =\
\delta^{(8)} \left(\sum_{i=1}^n \lambda_{i a}
\eta^A_i \right)\
{1 \over \prod_{i=1}^n \vev{i~i+1}} \ .
\label{nair}
\ee
Here $\eta^A_i$ are anticommuting variables and $A=1,2,3,4$ is an R-symmetry $SU(4)$ index.
The Grassmann-valued delta function is defined as follows,
\EQ{\label{delta8}
\delta^{(8)} \left(\sum_{i=1}^n \lambda_{i a}
\eta^A_i \right) \equiv \ \prod_{A=1}^4 \, \hf
\left(\sum_{i=1}^n \lambda_{i }^a \eta^A_i \right)
\left(\sum_{i=1}^n \lambda_{i a} \eta^A_i \right) \ .
}
By Taylor expanding \eqref{nair} in powers of $\eta_i$, one can identify
each term in the expansion with a particular tree-level analytic amplitude
in the $\cN=4$ theory.  Each factor of $(\eta_i)^k$ for $k=0,\ldots,4$ is interpreted as
the $i^{\rm th}$ particle with helicity $h_i=1-{k\over 2}$.
This implies that helicities take values,
$\{1,{1\over 2},0,-{1\over 2},-1\},$ which precisely correspond to the particles of
the $\cN=4$ supermultiplet,
$\{g^-,\Lambda^{-}_A,\phi^{AB},\Lambda^{A+},g^+\}.$

Actually, it is straightforward to
associate a single power of
$\eta$ with all component fields in $\cN=4$ \cite{GK}.
This can be done if we package all on-shell fields of $\cN=4$
into a superfield or equivalently by applying the following rules
\SP{ \label{nrules}
g^{-}_i\ \sim\ \eta_i^1 \eta_i^2 \eta_i^3 \eta_i^4 \ , \quad
&\phi^{AB}_i \ \sim\  \eta_i^A \eta_i^B \ , \qquad
\Lambda^{A+}_i \ \sim\  \eta_i^A \ , \qquad \qquad
g^{+}_i\ \sim\  1  \ , \\
\Lambda^{-}_{1} \ \sim\  -\,\eta_i^2 \eta_i^3 \eta_i^4 \ , \quad
&\Lambda^{-}_{2\,i} \ \sim\  \,\eta_i^1 \eta_i^3 \eta_i^4 \ , \quad
\Lambda^{-}_{3\,i} \ \sim\  -\,\eta_i^1 \eta_i^2 \eta_i^4 \ ,
\quad
\Lambda^{-}_{4\,i} \ \sim\  \,\eta_i^1 \eta_i^2 \eta_i^3 \ .
}


The analytic amplitudes are of degree-8 and they are the elementary building
blocks of the scalar graph approach. The next-to-minimal case are
the amplitudes of degree-12 in $\eta$ which are obtained by connecting
two analytic vertices \cite{Nair} with a scalar propagator $1/P^2.$
Each analytic vertex
contributes 8 $\eta$'s while a propagator removes 4 $\eta$'s.
In fact, any $n$-point amplitude is characterised by a degree $8,12, 16, \ldots, (4n-8)$
and can be obtained from scalar diagrams with $1,2,3, \ldots$ analytic
supervertices.

We now consider the fist non-minimal case of a diagram with two analytic supervertices
\eqref{nair} connected by a single scalar propagator.
The diagram is depicted in Figure 6.
The right supervertex has $n_1$ lines while the left one has
$n_2$ lines in total, such that resulting amplitude $A_n$ has
$n=n_1+n_2-2$ external lines.
Suppressing summations over the distribution of $n_1$ and $n_2$
between the two supervertices, one can write down the expression for the amplitude of Figure 6 as:
\SP{ \label{itern1}
A_n =\ &{1 \over \prod_{l=1}^n\ \vev{l~l+1}}\
{1 \over P_I^2} \
{\vev{j_1~j_1+1} \vev{i_1~i_1+1} \over \vev{i_1~\bar{I}}\vev{\bar{I}~j_1+1}
\vev{j_1~I}\vev{I~i_1+1}} \\
&\times \int \prod_{A=1}^4 d \eta^A_I \
\delta^{(8)} \left(\lambda_{\bar{I}a}\eta_I^A +
\sum_{l_2 \neq \bar{I}}^{n_2} \lambda_{l_2 a}
\eta^A_{l_2} \right) \
\delta^{(8)} \left(\lambda_{Ia}\eta_I^A + \sum_{l_1 \neq I}^{n_1} \lambda_{l_1 a}
\eta^A_{l_1} \right)
 \ .
}
The summations in the delta-functions arguments run over the
$n_1-1$ external lines for right vertex, and $n_2-1$ external lines for the left one.
The integration over the 4 $\eta$'s of the internal line arises because the two separate (unconnected) vertices in Figure 6
would have $n_1+ n_2$ lines and, hence, $n_1+ n_2$ different
$\eta$'s  while in the final amplitude there must be have $n=n_1+n_2-2$ $\eta$'s.
This is achieved
by letting
\EQ{ \eta_{\bar{I}}^A = \ \eta_{I}^A \ , }
and integrating over $d^4 \eta_I.$ since the $I$ and $\bar{I}$ internal
lines are connected by the propagator.
The off-shell continuation of the internal spinors is defined
as usual through,
\EQ{
\lambda_{I a} = \ \sum_{l_1 \neq I}^{n_1}\,
p_{l_1\, a \dot{a}}\ \zeta^{\dot a} = \ - \lambda_{\bar{I}a} \ .
\label{rlsls}}

The final step is to integrate out four $\eta_I$'s by rearranging the arguments of
the delta functions by use of $\int \delta(f_2)\delta(f_1) =\int \delta(f_1+f_2)\delta(f_1)$.
Notice that the sum of two arguments, $f_1+f_2,$ does not depend on $\eta_I.$

The final result of which we will make use in the following Section is
\EQ{\label{finrest}
A_n =\ {1 \over \prod_{l=1}^n\ \vev{l~l+1}}\
\delta^{(8)} \left(\sum_{i=1}^n \lambda_{i a} \eta^A_i \right)\
\prod_{A=1}^4\left(\sum_{l_1 \neq I}^{n_1}\, \vev{I~l_1} \eta^A_{l_1} \right) \
{1\over D} \ ,
}
and $D$ is the same as \eqref{Ddef} used in sections 3 and 4,
\EQ{\label{Pdef}
{1\over D}\ = \ {1 \over P_I^2} \
{\vev{j_1~j_1+1} \vev{i_1~i_1+1} \over \vev{i_1~I}\vev{I~j_1+1}
\vev{j_1~I}\vev{I~i_1+1}} \ .
}
As mentioned above, there are 12 $\eta$'s in the superamplitude \eqref{finrest}.
All component amplitudes of degree-12 can be obtained by expanding this expression
in powers of $\eta$.

\subsection{Multi-soft limit of 2 scalars or 2 fermions forming a singlet and $m-2$ positive helicity gluons}

\begin{figure}[h]
\label{fig3}
\begin{center}
{\scalebox{0.89}{
\includegraphics{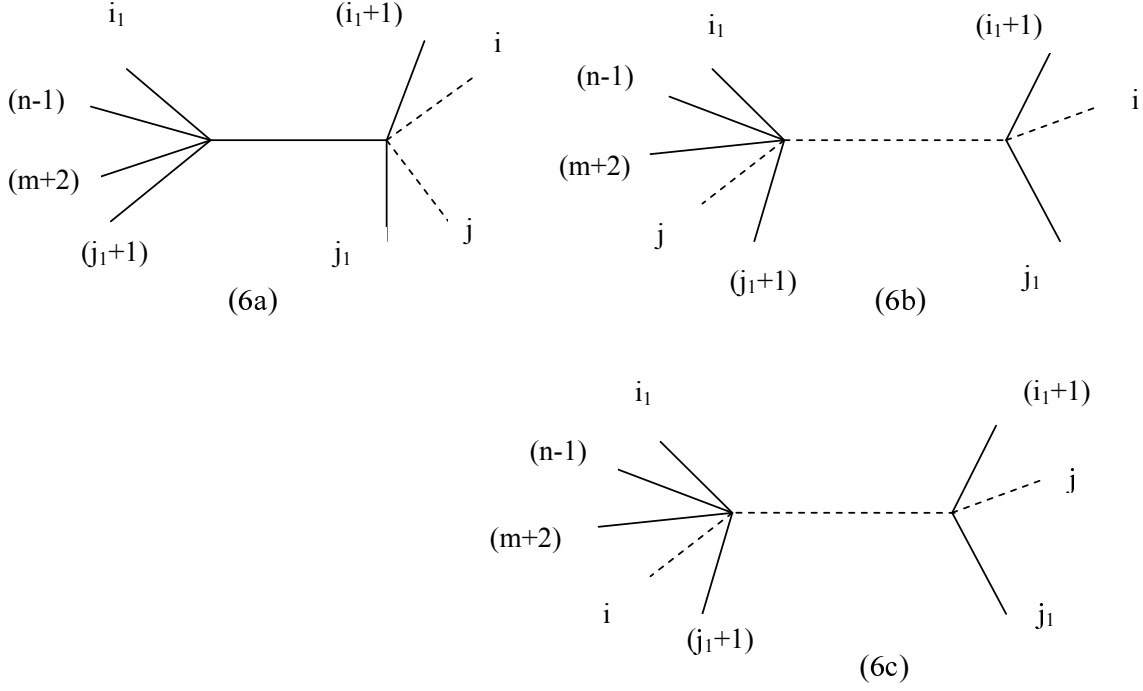}}
}
\end{center}
\caption{\it Diagrams relevant for the multi-soft limit of m particles 2 of which are scalars or fermions forming a singlet under $SU(4)$ and sitting at positions $i$ and $j$.
Notice that $i$ and $j$ is not needed to be adjacent.
The class of diagrams on the left are the only which contribute. Diagrams (6b) and (6c) can not contribute because the right supervertex is not of degree-8. This is a consequence of
the fact that all particles of the right vertex except the dashed ones are taken to be positive helicity gluons and as such do not give any $\eta$ dependence. }
\end{figure}
Based on the discussion of the last Section it is now straightforward to obtain compact expressions for
the leading behaviour of the multi-soft limits in two interesting cases. The first case concerns the  multi-soft limit of
two fermions forming a singlet and $m-2$ positive helicity gluons, while the second one is that
of two scalars forming a singlet and $m-2$ positive helicity gluons in $N=4$ SYM.
The case of two scalars or the two fermions that do not form a singlet under $SU(4)$ and no gluons had been treated recently in
\cite{Volovich:2015yoa}.

The fist step consists in identifying the relevant for the soft-limit diagrams. These are shown
in Figure 6. The two scalars or fermions sit at the positions $i$ and $j$ while all other soft particles from 1 to $m$
are taken to be positive helicity gluons. As in the purely gluonic case the diagrams where two or more hard particles
are attached to the same analytic supervertex with the negative helicity soft particle do not contribute in the leading order.
One can now make use of \eqref{finrest} for each case separately. What is important is that only the diagrams of (6a) where
both the fermions or the scalars are on the right supervertex can contribute. The diagrams (6b) and (6c) with one fermion or scalar per vertex
are not present since the supervertex on the right  can never be of degree 8 in the $\eta$ expansion as an analytic supervertex should be.
This is so because all other external particles are taken to be positive helicity gluons which do not give any $\eta$ dependence.
Notice that this is not the case for the diagram (6a) since there the right vertex is precisely of degree-8. Four $\eta$s come from the two external
fermions or scalars and another 4
from the negative helicity gluon propagator.

Consequently, what one has to do is to expand the product in \eqref{finrest} and keep the term proportional to $\eta^{A=1}_i\eta^{A=2}_i\eta^{A=3}_j\eta^{A=4}_j$
for the case of the two scalars sitting at positions $i$ and $j$, i.e. $\phi^{12}_i$ and $\phi^{34}_j$.
For the case of two fermions one should keep the term proportional either to $\eta^{A=1}_i\eta^{A=2}_j\eta^{A=3}_j\eta^{A=4}_j$ when the order of the fermions is
$\Lambda^{A=1+}_i \Lambda^{-}_{A=1j}$ or to $\eta^{A=2}_i\eta^{A=3}_i\eta^{A=4}_i\eta^{A=1}_j$ when the order of the fermions is
$\Lambda^{-}_{A=1 i} \Lambda^{A=1+}_{j}$.
The computations resembles the gluonic case so we will omit the details.
In all cases the final result takes the form
\SP{ \label{analgen}
A_n\rightarrow S_m(i,j)A_{n-m},
}
where the multi-soft factor can be written in a unified form as
\SP{ \label{ijgen}
S_m(i,j) ={(-1)^{1-\theta(r-s)} \vev{n~m+1}\over \prod_{q=n}^{m} \vev{q~q+1}}
\sum_{i_1={n-1}}^{ i-1\,\,\,\,\,'} \sum_{j_1=j}^{ m+1\,\,\,\,\,'}{\vev{i^-|{p\!\!\!/}_{i_1+1,j_1}| \zeta^-}^r
\vev{j^-|{p\!\!\!/}_{i_1+1,j_1}| \zeta^-}^s\over D},
}
with $D$ being precisely the quantity of \eqref{Ddef} and $\theta(x)$ being the step function, $\theta(x)=1$ for $x\geq 0$ while $\theta(x)=0$ for $x< 0$ .
When $r=s=2$ we have the case of the two scalars while the case $r=3, \,\,\, s=1$ corresponds to $\Lambda^{-}_{A=1 i} \Lambda^{A=1+}_{j}$
while the case $r=1, \,\,\, s=3$ corresponds to $\Lambda^{A=1+}_i \Lambda^{-}_{A=1j}$.
As in Section 2 the prime over the sums is there to remind us that performing the sum we should omit the term with $i_1=n-1$ and $j_1=m+1$.
As above the result \eqref{ijgen} is independent of $\zeta$ which can be chosen to be $\zeta^{\dot \alpha }={\tilde \lambda}_1^{\dot \alpha }$.

One important comment is in order. By counting the number of inverse power of $\delta$ it is easy to see that
in all cases the soft factor diverges like $1/\delta^m$, which is of the same order as in the case of the multi-soft limit of
one negative helicity gluon and $m-1$ gluons of positive helicity. This is quite natural from the point of view of the construction
based on the analytic supervertex since all these cases fall in the same category where four $\eta$s are extracted from the external legs of the
right supervertex.

\subsection{Double-soft limit of 2 scalars forming a singlet}

\begin{figure}[h]
\label{fig3}
\begin{center}
{\scalebox{0.89}{
\includegraphics{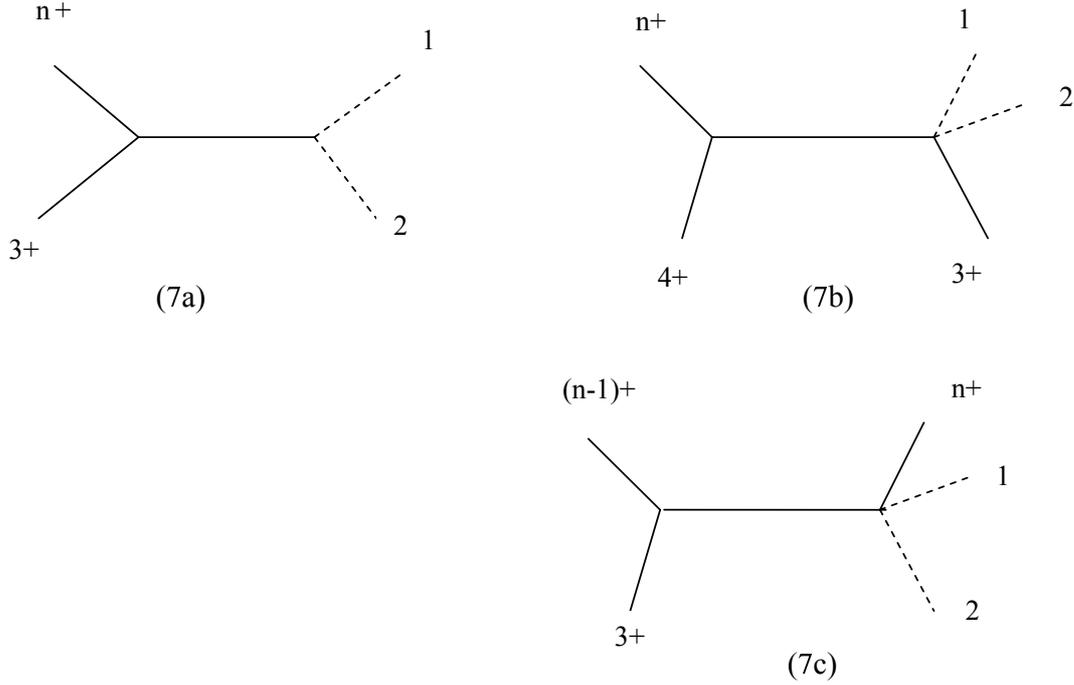}}
}
\end{center}
\caption{\it Diagrams contributing in the double-soft limit of 2 scalars or 2 fermions in a singlet configuration. Notice that as in Figure 6 both particles should be on the
same vertex. }
\end{figure}
In this Section, we consider the special case of the double-soft limit of two scalars which transform as a singlet
under the $SU(4)$ R-symmetry of $N=4$ SYM. This particular case has been considered in \cite{Klose:2015xoa}.
Here, we will derive a simpler result that is in agreement with that of \cite{Klose:2015xoa} (eq. (109)).
As discussed at the end of the last Section, this double-soft factor behaves as $1/\delta^2$.

The result can be read from \eqref{ijgen}. For the sake of completeness we draw the set of contributing diagrams in Figure 7.
There are three contributions from diagrams coming from (7a), (7b) and (7c). Let us start by writing down the contribution of (7c).
By choosing the reference spinor to be $\zeta^{\dot \alpha }={\tilde \lambda}_1^{\dot \alpha }$ we get from \eqref{ijgen}
\SP{ \label{7c}
S_2^{(7c)}(\phi^{12}_1,\phi^{34}_2)={ \vev{n~3}\over \vev{n~1} \vev{1~2} \vev{2~3}}
{\vev{1^-|{p\!\!\!/}_{n}| 1^-}^2
\vev{2^-|{p\!\!\!/}_{n}| 1^-}^2 \vev{n-1~n}\vev{2~3}
\over 2 p_n \cdot q_{12}
\vev{2^-|{p\!\!\!/}_{n}| 1^-}\vev{3^-|{p\!\!\!/}_{n}| 1^-}\vev{n-1^-|{p\!\!\!/}_{n}| 1^-}\vev{n^-|{p\!\!\!/}_{12}| 1^-}}
}
After a bit of algebra this expression simplifies to
\SP{ \label{7cfin}
S_2^{(7c)}(\phi^{12}_1,\phi^{34}_2)={\vev{n~1}[n1]\over 2 p_n \cdot q_{12} \vev{1~2}[21]}.
}
Similarly, the diagram (7b) gives
\SP{ \label{7b}
S_2^{(7b)}(\phi^{12}_1,\phi^{34}_2)={\vev{1~3}[31]\over 2 p_3 \cdot q_{12} \vev{1~2}[21]}.
}
Finally, our choice of the reference spinor $\zeta^{\dot \alpha }={\tilde \lambda}_1^{\dot \alpha }$
gives zero for the diagram of (7a) since $S_n^{(7a)}(\phi^{12}_1,\phi^{34}_2)\sim \vev{2^-|{q\!\!\!/}_{12}| 1^-}=0$.
Overall we get
\SP{ \label{7scalars}
S_2(\phi^{12}_1,\phi^{34}_2)=\frac{1}{\vev{1~2}[21]}\Big( {\vev{n~1}[n1]\over 2 p_n \cdot q_{12} } +
{\vev{1~3}[31]\over 2 p_3 \cdot q_{12} } \Big).
}
One can check that \eqref{7scalars} is a simpler representation of the double-soft factor of two scalars in a singlet configuration first obtained in
\cite{Klose:2015xoa}.
As advertised above, this result scales as $1/\delta^2$ in the double-soft limit.

\subsection{Double-soft limit of 2 fermions forming a singlet}

In a similar fashion it is straightforward to obtain the following result for the case of two soft fermions.
The relevant diagram are again those of Figure 7 with the only difference that now the dashed line denote fermions.
Our choice of the reference spinor will again make the diagram of (7a) zero while the sum of the other two diagrams
sum  up to
\SP{ \label{7fermions}
S_2(\Lambda^{A=1 +}_1\Lambda^{-}_{A=1\,\,2})=-\frac{1}{\vev{1~2}[21]}\Big( {\vev{n~2}[n1]\over 2 p_n \cdot q_{12} } +
{\vev{2~3}[31]\over 2 p_3 \cdot q_{12} } \Big).
}
As in the case of two scalars this result scales as $1/\delta^2$ in the double-soft limit.
This should be contrasted to the case of two fermions transforming non-trivially under the R-symmetry $SU(4)$.
In the latter case the amplitude behaves like $1/\delta$ in the double-soft limit \cite{Volovich:2015yoa}.

\subsection{Multi-soft limit of (2 fermions +  1 scalar) and (4 fermions) forming a singlet}

We close this Section by considering two last cases which can be easily extracted from the superamplitude  \eqref{finrest}.
The first one is that of the triple-soft limit of a scalar $\phi^{12}_1$ sitting at position 1 and two positive helicity fermions
at positions 2 and 3, that is $\Lambda^{3+}_2$ and $\Lambda^{4+}_3$.
The diagrams which contribute are similar to those of Figure 7 with the only difference that there will be now three dashed lines instead of two
denoting the 3 soft particles. Sticking with the same choice of the reference spinor as above and after a bit of spinor algebra we get
\SP{ \label{2f1s}
S_3(\phi^{12}_1,\Lambda^{A=3+}_2,\Lambda^{A=4+}_3)=\frac{1}{\vev{1~2}\vev{2~3}}\Big(\frac{\vev{1~n}\vev{2~n}[n1]}{2 p_n \cdot q_{13} \vev{n^-|{q\!\!\!/}_{2,3}| 1^-}}-\frac{\vev{1~4}\vev{2~4}[41]}{2 p_4 \cdot q_{13} \vev{4^-|{q\!\!\!/}_{2,3}| 1^-}}+ \\   \frac{\vev{1^-|{q\!\!\!/}_{2,3}| 1^-}\vev{2~3}[31]\vev{n~4}}{q_{13}^2\vev{4^-|{P\!\!\!/}_{2,3}| 1^-}\vev{n^-|{P\!\!\!/}_{2,3}| 1^-}}\Big).
}                                                                                                                                                                                            Similarly, for the case of the simultaneous soft limit of 4 positive helicity fermions we obtain
\SP{ \label{4f}
S_4(\Lambda^{1+}_1,\Lambda^{2+}_2,\Lambda^{3+}_3,\Lambda^{4+}_4)=\frac{1}{\vev{1~2}\vev{2~3}\vev{3~4}}\Big(\frac{\vev{2~n}\vev{3~n}[n1]}{2 p_n \cdot q_{14} \vev{n^-|{q\!\!\!/}_{2,4}| 1^-}}-\frac{\vev{2~5}\vev{3~5}[51]}{2 p_5 \cdot q_{14} \vev{5^-|{q\!\!\!/}_{2,4}| 1^-}}+ \\   \frac{\vev{2^-|{q\!\!\!/}_{3,4}| 1^-}\vev{3^-|{q\!\!\!/}_2+{q\!\!\!/}_4| 1^-}\vev{n~5}}{q_{14}^2\vev{5^-|{q\!\!\!/}_{2,4}| 1^-}\vev{n^-|{q\!\!\!/}_{2,4}| 1^-}}\Big).
}
Notice that the 4-fermion and  1 scalar with 2 fermions soft factors scale as $1/\delta^4$ and $1/\delta^3$ respectively, in the multi-soft limit of \eqref{soft-lim}.
Finally, it is straightforward to write down the result for the multi-soft limit of 4 positive helicity fermions and any number of positive helicity gluons, as well as for
the case of  1 scalar with 2 fermions and any number of positive helicity gluons along the lines of Section 3.2.

\section{Conclusions}

In this present work we employ the MHV technique to show that scattering amplitudes with any number of consecutive soft particles behave universally in the multi-soft limit when all particles go soft simultaneously.
In particular, we have shown how one can use the MHV diagrams to calculate the leading singularity of the simultaneous multi-soft limit of any number of consecutive particles.
After identifying the diagrams which give the leading contribution we give the general rules from which one can immediately write down compact expressions for $m$ gluons, $k$ of which are
negative helicity ones. As an example we explicitly give the expressions for the cases  of $k$ being equal to 1 and 2. In all cases the result takes the form of
the multi-soft factor of m positive helicity gluons times a function depending on the negative helicity ones.

Subsequently, we proceed to consider the case of amplitudes in $\cN =4$ super Yang Mills theory. In this case, the scalar graph method has as building blocks the  $\cN =4$ analytic supervertices. Using this technique we obtain the multi-soft factor in the limit where 2 scalars or 2 fermions forming a singlet
and $m-2$ positive helicity gluons become soft simultaneously. The double-soft limit of 2 scalars or 2 fermions forming a singlet gives an amplitude whose
leading divergence is $1/\delta^2$ and not $1/\delta$ as in the case of 2 scalars or 2 fermions not forming a singlet under $SU(4)$.
As a bonus of the construction based on the analytic supervertices we also obtain expressions for the triple-soft limit of 1 scalar and 2 positive helicity fermions, as well as
for the quadrapole-soft limit of 4 positive helicity fermions, in a singlet configuration. In all the cases we have considered the
amplitude has a leading divergence of $1/\delta^m$ in the soft limit (2.1).
Finally, we think that it would be interesting to extend the present method to extract simple expressions for the subleading terms of the multi-soft factors considered in this
article.
Furthermore, the multi-soft expressions obtained in this work could be a more efficient starting point for reconstructing amplitudes by the inverse soft limit.

\bigskip
\bigskip

\centerline{\bf Acknowledgements}

We thank George Savvidy for illuminating discussions.

\bigskip



\bigskip




\begin{thebibliography}{99}

\bibitem{soft-theor}
F.~Low,
Phys. Rev. 110, 974 (1958)
\\
S.~Weinberg,
Phys. Rev. 135, B1049 (1964)

\bibitem{sublead}
  F.~Cachazo and A.~Strominger,
  arXiv:1404.4091 [hep-th].
  E.~Casali,
  JHEP {\bf 1408}, 077 (2014)
  [arXiv:1404.5551 [hep-th]].
\bibitem{Adler}
S.~L.~Adler,
Phys. Rev. 137, B1022 (1965)

\bibitem{Kaplan}
  N.~Arkani-Hamed, F.~Cachazo and J.~Kaplan,
  JHEP {\bf 1009}, 016 (2010)
  [arXiv:0808.1446 [hep-th]].

\bibitem{Strominger-grav}
  T.~He, P.~Mitra and A.~Strominger,
  arXiv:1503.02663 [hep-th].
  D.~Kapec, V.~Lysov, S.~Pasterski and A.~Strominger,
  JHEP {\bf 1408}, 058 (2014)
  [arXiv:1406.3312 [hep-th]].
  T.~He, V.~Lysov, P.~Mitra and A.~Strominger,
  arXiv:1401.7026 [hep-th].

\bibitem{BMS}
  H.~Bondi, M.~G.~J.~van der Burg and A.~W.~K.~Metzner,
  Proc.\ Roy.\ Soc.\ Lond.\ A {\bf 269}, 21 (1962).
  R.~K.~Sachs,
  Proc.\ Roy.\ Soc.\ Lond.\ A {\bf 270}, 103 (1962).


\bibitem{Palatrussardi}
  G.~Barnich and C.~Troessaert,
  Phys.\ Rev.\ Lett.\  {\bf 105}, 111103 (2010)
  [arXiv:0909.2617 [gr-qc]];
  PoS CNCFG {\bf 2010}, 010 (2010)
  [arXiv:1102.4632 [gr-qc]];
  JHEP {\bf 1112}, 105 (2011)
  [arXiv:1106.0213 [hep-th]].

\bibitem{3-point}
  N.~Beisert, C.~Ahn, L.~F.~Alday, Z.~Bajnok, J.~M.~Drummond, L.~Freyhult, N.~Gromov and R.~A.~Janik {\it et al.},
  Lett.\ Math.\ Phys.\  {\bf 99}, 3 (2012)
  [arXiv:1012.3982 [hep-th]].
  G.~Georgiou,
  JHEP {\bf 1102}, 046 (2011)
  [arXiv:1011.5181 [hep-th]].
  G.~Georgiou,
  JHEP {\bf 1109}, 132 (2011)
  [arXiv:1107.1850 [hep-th]].
  G.~Georgiou, V.~Gili, A.~Grossardt and J.~Plefka,
  JHEP {\bf 1204}, 038 (2012)
  [arXiv:1201.0992 [hep-th]].
  C.~Durnford, G.~Georgiou and V.~V.~Khoze,
  JHEP {\bf 0609}, 005 (2006)
  [hep-th/0606111].
  C.~S.~Chu, G.~Georgiou and V.~V.~Khoze,
  JHEP {\bf 0611}, 093 (2006)
  [hep-th/0606220].

\bibitem{rev-amp}

L.~J.~Dixon,
hep-ph/9601359.
  H.~Elvang and Y.~t.~Huang,
  arXiv:1308.1697 [hep-th].
  J.~M.~Henn and J.~C.~Plefka,
``Scattering Amplitudes in Gauge Theories,''
Lect. Notes Phys. 883, 1 (2014)



\bibitem{Klose:2015xoa}
  T.~Klose, T.~McLoughlin, D.~Nandan, J.~Plefka and G.~Travaglini,
  arXiv:1504.05558 [hep-th].

\bibitem{Volovich:2015yoa}
  A.~Volovich, C.~Wen and M.~Zlotnikov,
  arXiv:1504.05559 [hep-th].

\bibitem{Chen:2014xoa}
  W.~M.~Chen, Y.~t.~Huang and C.~Wen,
  arXiv:1412.1809 [hep-th].

\bibitem{soft-loops}
  Z.~Bern, S.~Davies and J.~Nohle,
  Phys.\ Rev.\ D {\bf 90}, no. 8, 085015 (2014)
  [arXiv:1405.1015 [hep-th]].
    S.~He, Y.~t.~Huang and C.~Wen,
  JHEP {\bf 1412}, 115 (2014)
  [arXiv:1405.1410 [hep-th]].
  F.~Cachazo and E.~Y.~Yuan,
  arXiv:1405.3413 [hep-th].
  M.~Bianchi, S.~He, Y.~t.~Huang and C.~Wen,
  arXiv:1406.5155 [hep-th].
  J.~Broedel, M.~de Leeuw, J.~Plefka and M.~Rosso,
  Phys.\ Lett.\ B {\bf 746}, 293 (2015)
  [arXiv:1411.2230 [hep-th]].

\bibitem{otherdim}
  B.~U.~W.~Schwab and A.~Volovich,
  Phys.\ Rev.\ Lett.\  {\bf 113}, no. 10, 101601 (2014)
  [arXiv:1404.7749 [hep-th]].
  N.~Afkhami-Jeddi,
  arXiv:1405.3533 [hep-th].
    M.~Zlotnikov,
  JHEP {\bf 1410}, 148 (2014)
  [arXiv:1407.5936 [hep-th]].
  C.~Kalousios and F.~Rojas,
  JHEP {\bf 1501}, 107 (2015)
  [arXiv:1407.5982 [hep-th]].

  \bibitem{rest}
  A.~J.~Larkoski,
  Phys.\ Rev.\ D {\bf 90}, 087701 (2014)
  [arXiv:1405.2346 [hep-th]];
  F.~Cachazo and E.~Y.~Yuan,
  arXiv:1405.3413 [hep-th].
  J.~Broedel, M.~de Leeuw, J.~Plefka and M.~Rosso,
  Phys.\ Rev.\ D {\bf 90}, 065024 (2014)
  [arXiv:1406.6574 [hep-th]].
  Z.~Bern, S.~Davies, P.~Di Vecchia and J.~Nohle,
  Phys.\ Rev.\ D {\bf 90}, no. 8, 084035 (2014)
  [arXiv:1406.6987 [hep-th]].
  C.~D.~White,
  Phys.\ Lett.\ B {\bf 737}, 216 (2014)
  [arXiv:1406.7184 [hep-th]].
  M.~Campiglia and A.~Laddha,
  Phys.\ Rev.\ D {\bf 90}, no. 12, 124028 (2014)
  [arXiv:1408.2228 [hep-th]].
  Y.~J.~Du, B.~Feng, C.~H.~Fu and Y.~Wang,
  JHEP {\bf 1411}, 090 (2014)
  [arXiv:1408.4179 [hep-th]].
  H.~Luo, P.~Mastrolia and W.~J.~Torres Bobadilla,
  Phys.\ Rev.\ D {\bf 91}, no. 6, 065018 (2015)
  [arXiv:1411.1669 [hep-th]].
  A.~S.~Vera and M.~A.~Vazquez-Mozo,
  JHEP {\bf 1503}, 070 (2015)
  [arXiv:1412.3699 [hep-th]].
  A.~J.~Larkoski, D.~Neill and I.~W.~Stewart,
  arXiv:1412.3108 [hep-th].
  M.~Campiglia and A.~Laddha,
  arXiv:1502.02318 [hep-th].



\bibitem{BCFW}
R.~Britto, F.~Cachazo and B.~Feng,
  Nucl.\ Phys.\ B {\bf 715}, 499 (2005)
  [hep-th/0412308].
R.~Britto, F.~Cachazo, B.~Feng and E.~Witten,
  Phys.\ Rev.\ Lett.\  {\bf 94}, 181602 (2005)
  [hep-th/0501052];


\bibitem{CHY}
  F.~Cachazo, S.~He and E.~Y.~Yuan,
  Phys.\ Rev.\ Lett.\  {\bf 113}, no. 17, 171601 (2014)
  [arXiv:1307.2199 [hep-th]].

\bibitem{CSW}
  F.~Cachazo, P.~Svrcek and E.~Witten,
  JHEP {\bf 0409}, 006 (2004)
  [hep-th/0403047].

\bibitem{Witten}
  E.~Witten,
  Commun.\ Math.\ Phys.\  {\bf 252}, 189 (2004)
  [hep-th/0312171].

\bibitem{GK}
G.~Georgiou and V.~V.~Khoze,
JHEP {\bf 0405} (2004) 070,
hep-th/0404072.

\bibitem{GGK}
  G.~Georgiou, E.~W.~N.~Glover and V.~V.~Khoze,
  JHEP {\bf 0407}, 048 (2004)
  [hep-th/0407027].

\bibitem{MHV-loops}
  A.~Brandhuber, B.~J.~Spence and G.~Travaglini,
  Nucl.\ Phys.\ B {\bf 706}, 150 (2005)
  [hep-th/0407214].
  J.~Bedford, A.~Brandhuber, B.~J.~Spence and G.~Travaglini,
  Nucl.\ Phys.\ B {\bf 706}, 100 (2005)
  [hep-th/0410280].
  J.~Bedford, A.~Brandhuber, B.~J.~Spence and G.~Travaglini,
  Nucl.\ Phys.\ B {\bf 712}, 59 (2005)
  [hep-th/0412108].
  A.~Brandhuber, B.~Spence and G.~Travaglini,
  JHEP {\bf 0601}, 142 (2006)
  [hep-th/0510253].
  C.~Quigley and M.~Rozali,
  JHEP {\bf 0501}, 053 (2005)
  [hep-th/0410278].
\bibitem{me-wilson}
  L.~F.~Alday and J.~M.~Maldacena,
  JHEP {\bf 0706}, 064 (2007)
  [arXiv:0705.0303 [hep-th]].
  J.~M.~Drummond, J.~Henn, G.~P.~Korchemsky and E.~Sokatchev,
  Nucl.\ Phys.\ B {\bf 795}, 52 (2008)
  [arXiv:0709.2368 [hep-th]].
  A.~Brandhuber, P.~Heslop and G.~Travaglini,
  Nucl.\ Phys.\ B {\bf 794}, 231 (2008)
  [arXiv:0707.1153 [hep-th]].
  G.~Georgiou,
  JHEP {\bf 0909}, 021 (2009)
  [arXiv:0904.4675 [hep-th]].

\bibitem{CSW2}
  F.~Cachazo, P.~Svrcek and E.~Witten,
  JHEP {\bf 0410}, 074 (2004)
  [hep-th/0406177].


\bibitem{Nair} V.~P.~Nair,
``A Current Algebra For Some Gauge Theory Amplitudes,''
Phys.\ Lett.\ B {\bf 214} (1988) 215.


\end{thebibliography}
\end{document}